\documentclass[11pt]{article}
\usepackage[letterpaper, margin=1in]{geometry}
\usepackage{hyperref}

\usepackage{setspace}
\usepackage{latexsym,amssymb,amsmath,amsthm,graphicx,float,mathabx,authblk}

\usepackage{caption}
\usepackage{subcaption}
\usepackage{hyperref}

\usepackage{nicefrac}
\usepackage{xcolor}
\usepackage{tikz}
\usetikzlibrary{shapes}
\usetikzlibrary{plotmarks}
\usetikzlibrary{decorations.pathreplacing}
\tikzset{
    cross/.pic = {
    \draw[rotate = 45] (-#1,0) -- (#1,0);
    \draw[rotate = 45] (0,-#1) -- (0, #1);
    }
}
\graphicspath{{./figs/}}

\usepackage[style=numeric-comp,
maxcitenames=2,
maxnames = 5,
firstinits=true,
uniquename=init,
sorting=nyt,
url=false,
isbn=false,
eprint=false,
texencoding=utf8,
bibencoding=utf8,
autocite=superscript,
backend=biber
]{biblatex}
\makeatletter
\newcommand\footnoteref[1]{\protected@xdef\@thefnmark{\ref{#1}}\@footnotemark}
\makeatother

\newtheorem{prop}{Proposition}

\newcommand{\argmax}{\mbox{argmax}}

\newcommand{\conj}{\overline}
\newcommand{\beq}{\begin{equation}}
\newcommand{\eeq}{\end{equation}}
\renewcommand{\tilde}{\widetilde}
\renewcommand{\hat}{\widehat}
\newcommand{\bit}{\begin{itemize}}
\newcommand{\eit}{\end{itemize}}
\newcommand{\ben}{\begin{enumerate}}
\newcommand{\een}{\end{enumerate}}
\newcommand{\bp}{\begin{pmatrix}}
\newcommand{\ep}{\end{pmatrix}}

\newtheorem{remark}[subsection]{Remark}

\title{Superresolution with the zero-phase imaging condition}
\author{Sarah Greer and Laurent Demanet}
\affil{Massachusetts Institute of Technology\\
         Cambridge, MA 02139, USA}

\let\svthefootnote\thefootnote
\newcommand\freefootnote[1]{%
  \let\thefootnote\relax%
  \footnotetext{#1}%
  \let\thefootnote\svthefootnote%
}

\begin{document}
\maketitle

\freefootnote{{ \it Email addresses: } \texttt{\href{mailto:sygreer@mit.edu}{sygreer@mit.edu}}, \texttt{\href{mailto:laurent@math.mit.edu}{laurent@math.mit.edu}}}

\begin{abstract}

Wave-based imaging techniques use wavefield data from receivers on the boundary of a domain to produce an image of the underlying structure in the domain of interest.
These images are defined by the imaging condition, which maps recorded data to their reflection points in the domain. 
In this paper, we introduce a nonlinear modification to the standard imaging condition that can produce images with resolutions greater than that ordinarily expected using the standard imaging condition. 
We show that the phase of the integrand of the imaging condition, in the Fourier domain, has a special significance in some settings that can be exploited to derive a super-resolved modification of the imaging condition.
Whereas standard imaging techniques can resolve features of a length scale of $\lambda$, our technique allows for resolution level $R < \lambda$, where the super-resolution factor (SRF) is typically $\lambda/R$.
We show that, in the presence of noise, $R  \sim \sigma$.

\end{abstract}

\section{Introduction}

Superresolution techniques allow for signal recovery at better resolutions than the diffraction limit of the governing problem may suggest. 
However, such recovery is delicate, inherently non-linear, and often not generalizable, as specific conditions must be satisfied for stable recovery \cite{donoho1992,mtsr}.
Various methods have been proposed for superresolution including subspace methods such as MUSIC \cite{music,musicliao}, maximum entropy techniques \cite{nbo}, matrix pencil methods \cite{matpen}, the superset method \cite{rlsr,ss2d} and optimization-based $\ell_1$ minimization \cite{Donoho2005,Fuchs2005,Lou2016}. 

The standard imaging condition of wave-based imaging, which is defined by the adjoint-state method \cite{asm,fori}, and often called reverse-time migration (RTM), is 
\begin{equation}
I(\mathbf{x}) = \sum_{s} \int_B u_{0s}(\mathbf{x},\omega) \conj{q_s(\mathbf{x},\omega)} \omega^2 d\omega\;.
\label{eq:ic}
\end{equation}
This maps the input forward ($u_{0s}(\mathbf{x}, \omega)$) and recorded adjoint ($q_s(\mathbf{x},\omega)$) wavefields to an output image of the underlying model, $I(\mathbf{x})$, where $\mathbf{x}$ is the domain of interest, $s$ is the source index, and $B$ is the frequency band. 
In this paper, we examine the superresolution question in imaging by applying a simple non-linear modification to this standard imaging condition.
Our method is limited to situations where the forward model is very structured; for example, in situations with a uniform medium with simple scatterers, while using the Born approximation with one-sided illumination. In these cases, there may be enough information to perform superresolution at the level of the imaging condition.

This is distinctly different from the concept of the deconvolution imaging condition, which performs a linear modification to the imaging condition that acts as an amplitude correction on the bandwidth of the image \cite{dic, radon}.
Rather, our method is a non-linear form of superresolution which corrects beyond the original bandwidth of the image. 

We show that, in the case of a single scatterer in a uniform background medium, in the Fourier domain, the derivative of the complex phase of $u_{0,s}(\mathbf{x}, \omega)q_s(\mathbf{x}, \omega)$ with respect to $\omega$ is a measure of distance to the scatterer location. 
Using this, one can create a mask to effectively filter out contributions in the imaging condition beyond the precise scatterer location.
This allows us to create a modified imaging condition, 
\begin{equation}
I(\mathbf{x}) = \sum_{s} \int_B \Gamma \left (u_{0s}(\mathbf{x},\omega) \conj{q_s(\mathbf{x},\omega)} \right ) \omega^2 d\omega\;,
\label{eq:zic}
\end{equation}
which is the same as Equation \ref{eq:ic} with the addition of an operator $\Gamma(\cdot)$ that examines the phase of the input and is explored further in the next section.
This theory is also expanded to cases where the scatterer is more elaborate than a single point, assuming for each time $t$, the wavefront of the incident field intersects the map of reflectors at a single location.
Numerical experiments and theory show that, for a single scatterer, the achieved resolution level is proportional to the noise level. 
Applications of this method may include geolocation and simple cases of ultrasound and radar imaging. 
\section{Theory}

In this  section, we describe in a simple case how the complex phases of the fields entering the imaging condition combine in a way that reveals the distance to the scatterer(s). As a result, we can perform a form of superresolution by restricting scatterer location in the image to within less than a wavelength of the true scatterer locations.

\subsection{Analysis of the phase}
We begin by looking at the simple case of a single scatterer located at $\mathbf{x} = \mathbf{x}^*$ in a medium with a uniquely valued 2-point traveltime $\tau(\mathbf{x},\mathbf{y})$ throughout the domain of interest. 
In a uniform medium, with wave velocity $c_0$, $\tau(\mathbf{x}, \mathbf{y}) = \nicefrac{|\mathbf{x} - \mathbf{y}|}{c_0}$; otherwise, $\tau$ obeys the eikonal equation $|\nabla_\mathbf{x} \tau(\mathbf{x}, \mathbf{y}) | = \nicefrac{1}{c_0(\mathbf{x})}$.
We assume a continuum of receivers on at least part of the boundary of the domain.

The simplest setting for imaging is the assumption of single scattering, where a forward field propagates through the domain, interacts with discontinuities, and is recorded at receivers on the boundary as singularly reflected waves. 
Using the Born approximation in the frequency domain, this forward field can be separated into two parts: the incident field, $u_{0,s}(\mathbf{x}, \omega)$, and the scattered field, $u_{1,s}(\mathbf{x}, \omega)$, with source index $s$.
Amplitudes aside, the incident field is 
\begin{equation}
u_{0,s}(\mathbf{x}, \omega) = G(\mathbf{x}_s, \mathbf{x}, \omega) w(\omega) \sim e^{i \omega \tau(\mathbf{x}_s, \mathbf{x})}\;,
\label{eq:inc}
\end{equation}
where a wave originates at the source location $\mathbf{x}_s$ and propagates through the domain of interest. 
Here, $G(\mathbf{x}_s, \mathbf{x}, \omega)$ is the Green's function, $w(\omega)$ is a wavelet, and 
the symbol $\sim$ refers to the fact that the oscillations of $u_{0, s}(\mathbf{x}, \omega)$ are, to leading order, determined by $e^{i\omega\tau(\mathbf{x},\mathbf{y})}$; i.e., we disregard a multiplicative amplitude factor that varies much more slowly with respect to $\omega$ \cite{witham}.
The scattered field contains primary reflections, where a wave originates at $\mathbf{x}_s$, reflects off of the scatterer at $\mathbf{x}^*$, and propagates through the domain: 
\begin{equation*}
u_{1,s}(\mathbf{x},\omega) \sim e^{i \omega \left [ \tau(\mathbf{x}_s, \mathbf{x}^*) + \tau(\mathbf{x}^*, \mathbf{x})\right ] } \;.
\end{equation*}
Through the adjoint-state method, the scattered field is embedded in the formulation of the adjoint field: for a single receiver $r$, the recorded scattered field is backpropagated through the domain of interest from its corresponding receiver location $\mathbf{x}_r$. 
Then the adjoint field for a single receiver $r$ is
\begin{equation*}
q_{r, s} (\mathbf{x}, \omega) \sim e^{i \omega \left [ \tau(\mathbf{x}_s, \mathbf{x}^*) + \tau(\mathbf{x}^*, \mathbf{x}_r) \right ] } e^{-i\omega\tau(\mathbf{x}_r, \mathbf{x}) }\;.
\end{equation*}
For receivers on the boundary of the domain of interest, the adjoint field becomes
\begin{equation*}
q_{s}(\mathbf{x}, \omega) \sim \sum_r q_{r, s} (\mathbf{x}, \omega)\;.
\end{equation*}
Then, in a frequency band $B$, the standard imaging condition in the Fourier domain is Equation \ref{eq:ic}.

The standard imaging condition (Equation \ref{eq:ic}) produces a strong response when the incident and adjoint fields kinematically coincide at singular features in the domain; 
i.e., when the complex phases of $u_{0, s}(\mathbf{x}, \omega)$ and $q_s(\mathbf{x}, \omega)$ mostly match, so that $u_{0, s}(\mathbf{x},\omega)\conj{q_s(\mathbf{x}, \omega)}$ has a strong constant (DC) component as a function of $\omega$. 
We are particularly interested in the case of a single scatterer: 
While the phase of the incident field has a clear interpretation, the phase of the adjoint field is obscured by the sum over receivers.
We wish to simplify the phase of the adjoint field at any given location so it can be mapped back to a single traveltime. 
In order to do this, we perform a stationary phase analysis of the adjoint field's $r$-dependent phase.

We first view our sum over receivers as the integral of a properly sampled smooth function, and introduce a smooth function $\chi(\mathbf{x}_r)$ to 
window out receivers near the edge of the receiver array to minimize acquisition geometry truncation artifacts. 
Then, we can reintroduce the adjoint field $q_s(\mathbf{x}, \omega)$ as 
\begin{equation*}
    q_s(\mathbf{x}, \omega) \simeq e^{i\omega \tau(\mathbf{x}_s, \mathbf{x}^*)}\int_{\mathbf{a}}^{\mathbf{b}} a(\mathbf{x}_r, \mathbf{x}, \omega)e^{i \omega \phi (\mathbf{x}_r)}  \chi(\mathbf{x}_r) dl(\mathbf{x}_r)\;,
\end{equation*}
where $a(\mathbf{x}_r, \mathbf{x}, \omega)$ is real and $\phi(\mathbf{x}_r) = \tau(\mathbf{x}^*, \mathbf{x}_r) - \tau(\mathbf{x}_r, \mathbf{x})$. 
This integral lends itself to stationary phase analysis. In the arclength parameterization, we are led to considering the tangential derivative 
\begin{equation*}
 \mathbf{t}(\mathbf{x}_r)\cdot \nabla_{\mathbf{x}_r} \phi(\mathbf{x}_r) \;,
\end{equation*}
where $\mathbf{t}(\mathbf{x}_r) $ is the direction tangential to the receiver array.
The stationary phase points
occur when $\mathbf{t}(\mathbf{x}_r) \cdot \nabla_{\mathbf{x}_r} \phi(\mathbf{x}_r) = 0$. 
From the analysis in Appendix \ref{ap:stphpt}, we find that this condition implies that $\mathbf{x}_r$ is on the ray $[\mathbf{x}, \mathbf{x}^*]$.
Since the $[\mathbf{x}, \mathbf{x}^*]$ ray intersects with the domain boundary at two points, there may be up to two such receivers on the ray; in our analysis we assume that there is only one receiver, called $\mathbf{x}_{\widetilde{r}}$, present on this ray. 
If more than one receiver exists on the ray linking $\mathbf{x}$ and $\mathbf{x}^*$, then without loss of generality, the receiver set can be broken up into two or more segments.

Then, to leading order from the point of view of oscillations \cite{ammse}, the adjoint field becomes 
\begin{equation*}
    q_s(\mathbf{x}, \omega) \sim e^{i \omega \left [ \tau(\mathbf{x}_s, \mathbf{x}^*) + \phi(\mathbf{x}_{\widetilde{r}}) \right ]}\;, 
\end{equation*}
where 
\begin{equation}
\phi(\mathbf{x}_{\widetilde{r}}) = \tau(\mathbf{x}^*, \mathbf{x}_{\widetilde{r}}) - \tau(\mathbf{x}_{\widetilde{r}}, \mathbf{x}) = \pm \tau(\mathbf{x}, \mathbf{x}^*)
\label{eq:phi}
\end{equation} 
from ray geometry. 
The plus sign is chosen when $\mathbf{x}_{\widetilde{r}}$ is on the $\mathbf{x}$ side of the ray, and the minus sign is chosen when $\mathbf{x}_{\widetilde{r}}$ is on the $\mathbf{x}^*$ side of the ray.
Then, our adjoint field can be simplified as 
\begin{align}
q_{s} (\mathbf{x}, \omega) & \sim e^{i \omega \left [ \tau(\mathbf{x}_s, \mathbf{x}^*) + \tau(\mathbf{x}^*, \mathbf{x}_{\widetilde{r}})  - \tau(\mathbf{x}_{\widetilde{r}}, \mathbf{x}) \right ]}\nonumber \\ 
 & \sim e^{i \omega \left [ \tau(\mathbf{x}_s, \mathbf{x}^*) \pm \tau(\mathbf{x}^*, \mathbf{x}) \right ]}\;,
\label{eq:adj}
\end{align}
and from Equations \ref{eq:inc}, \ref{eq:ic}, and \ref{eq:adj}, the phase of the integrand of the standard imaging condition (Equation \ref{eq:ic}) is
\begin{equation}
\label{eq:pic}
\psi_{\pm} \left (\mathbf{x} \right )  = \tau(\mathbf{x}_s, \mathbf{x}) - \tau(\mathbf{x}_s, \mathbf{x}^*) \mp \tau(\mathbf{x}^*, \mathbf{x}) \;, 
\end{equation} 
which no longer depends on the sum over receivers from the adjoint field. 
We can now use Equation \ref{eq:pic} as an indicator for the scatterer location---without restricting receiver placement, it is zero when $\mathbf{x} = \mathbf{x}^*$, and every point $\mathbf{x}$ on the ray connecting $\mathbf{x}_s$ and $\mathbf{x}^*$.
The former zero phase location indicates scatterer location, which we can utilize to superresolve a point scatterer.
The latter zero phase locations are from a transmission artifact in the imaging condition due to the presence of a receiver at $\mathbf{x}_r$ antipodal to the source $\mathbf{x}_s$, from which location of the scatterer is kinematically impossible \footnote{I.e., there is no receiver on the ray connecting $\mathbf{x}^*$ to $\mathbf{x}_s$ past $\mathbf{x}^*$, as shown in Figure \ref{fig:icpm}}. 
This can be alleviated by restricting receiver placement such that no receiver is placed antipodal to any source location. 
Then, we find:
\begin{prop}
Assuming there is no receiver antipodal to the source, the complex phase of the integrand of the imaging condition (Equation \ref{eq:ic}) is identically zero if and only if $\mathbf{x} = \mathbf{x^*}$.
\label{prop:antipodal}
\end{prop}
The proof of Proposition \ref{prop:antipodal} is in Appendix \ref{a:antipodal}.

\subsection{Zero-phase imaging condition} 
While the phase in Equation \ref{eq:pic} is zero at the location of the scatterer, 
a stronger condition is that the derivative of the phase of the integrand of the imaging condition with respect to $\omega$ is zero for all values of $\omega$ at the location of the scatterer, assuming no receivers are placed antipodal to the source. 
Additionally, we choose to consider $\frac{\partial \psi}{\partial \omega}$ and not $\psi$ as an indicator of the scatterer location since the former has the interpretation of distance to the scatterer in units of time, which can be seen from Equation \ref{eq:pic}.
With this, we can define the zero-phase imaging condition as
\begin{equation}
I(\mathbf{x}) = \sum_{s} \int_B \Gamma \left (u_{0s}(\mathbf{x},\omega) \conj{q_s(\mathbf{x},\omega)} \right ) \omega^2 d\omega\;,
\label{eq:zic}
\end{equation}
where if we decompose $u_{0, s}(\mathbf{x}, \omega)\conj{q_s(\mathbf{x}, \omega)}=r_s(\mathbf{x}, \omega)e^{i\theta_s(\mathbf{x},\omega)}$, then $\Gamma(\cdot)$ smoothly puts the complex amplitude to zero when $\frac{\partial \theta}{\partial \omega}(\mathbf{x}, \omega)$ departs from zero.
The $\Gamma(\cdot)$ operator can be implemented by applying a mask to the integrand of the imaging condition which only allows values where the derivative of the phase with respect to $\omega$ is small. 
In other words, we can define a value, $\tau_{_{RES}}$, which indicates the distance from the scatterer, in seconds, that is allowed to contribute to the produced image using the zero-phase imaging condition. 
That is,
 $\frac{\partial \theta}{\partial \omega} < \tau_{_{RES}}$, where the specific value of $\tau_{_{RES}}$ controls how superresolved the scatterer is. 

This can be implemented by first calculating the unwrapped phase of integrand of the imaging condition for each shot, then taking its derivative with respect to $\omega$, called $\frac{\partial \psi}{\partial \omega}$. Then, a $\tau_{_{RES}}$ threshhold is set, along with a taper size $s$, both in seconds. 
The mask $\gamma$ can then be created such that where $\frac{\partial \psi}{\partial \omega} > \tau_{_{RES}} + s$, $\gamma = 0$, 
where $\frac{\partial \psi}{\partial \omega} < \tau_{_{RES}} - s$, $\gamma = 1$, and where
$ \tau_{_{RES}} - s < \frac{\partial \psi}{\partial \omega} < \tau_{_{RES}} + s$, $\gamma$ linearly tapers between 0 and 1.
The $\gamma$ mask is calculated for each shot, and the product of all $\gamma$ masks is taken. Then, the $\Gamma(\cdot)$ operator involves applying the combined $\gamma$ mask to the integrand of the imaging condition.
This allows contributions to the imaging condition when $\frac{\partial \psi}{\partial \omega}$ is small, and avoids them otherwise.

\subsection{Noise analysis} 
We wish to understand the form of the perturbation of the integrand of Equation \ref{eq:ic} due to the presence of noise in the data, $d_{r, \omega, s}$, and its effect on the zero-phase imaging condition (Equation \ref{eq:zic}).
To do this, we start by adding noise to our observed data as
\begin{equation}
d_{r, \omega, s} = u_{1, s}(\mathbf{x}_r, \omega) + n_{r, \omega, s} \;,
\label{eq:noise}
\end{equation}
with a small perturbation $n_{r, \omega, s}$. 
The perturbed adjoint field is then
\begin{equation}
\widetilde{q}_s(\mathbf{x},\omega) = q_s(\mathbf{x}, \omega) + \sum_r \conj{G(\mathbf{x}_r, \mathbf{x}, \omega)} \; n_{r, \omega, s}\;,
\label{eq:paf}
\end{equation}
where complex conjugation indicates time reversal. 
All our fields are essentially bandlimited by means of a wavelet $w(\omega)$ (from Equation \ref{eq:inc}) that we assume to be, at the very least, integrable. 
Conequently, we quantify smallness of the noise as
\begin{equation*}
    \int \left | n_{r, \omega, s} \right | d\omega \leq \sigma \int \left | d_{r, \omega, s} \right | d\omega\;,
\end{equation*}
for some small, nondimensional $0 < \sigma < \nicefrac{1}{2}$.

\begin{figure}
\centering
\includegraphics{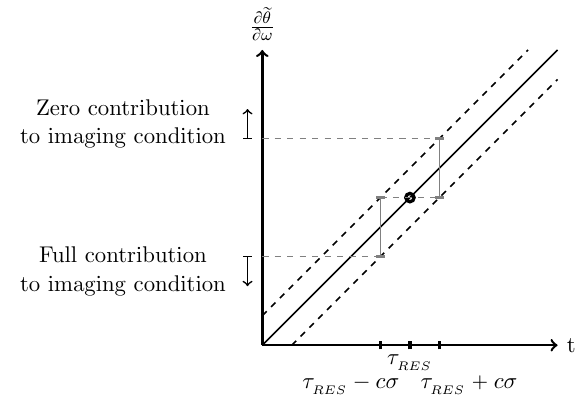}
\caption{The phase derivative as a function of $t$, which represents distance to the scatterer, located at $t = 0$. The two black dashed lines bound the possible values of $\frac{\partial \tilde{\theta}}{\partial \omega}$ in the presence of noise. From this, the operator $\Gamma(\cdot)$, which acts on the phase derivative $\frac{\partial \tilde{\theta}}{\partial \omega}$, allows full contribution to the image when $t < \tau_{_{RES}} -c\sigma$, and zero contribution to the image when $t > \tau_{_{RES}} + c\sigma$.
}
\label{fig:dtvt}
\end{figure}

In addition, it is reasonable to disregard the exceptional points $\mathbf{x}$ where $q_s(\mathbf{x}, \omega) = 0$ for some $\omega$ over the essential support of $w(\omega)$, which leads to the multiplicative model
\begin{equation}
    \label{eq:mulmod}
    \widetilde{q}_s(\mathbf{x}, \omega) = q_s(\mathbf{x}, \omega) \left (\rho(\mathbf{x}, \omega) e^{i \varphi(\mathbf{x}, \omega)} \right )\;,
\end{equation}
where we assume $|\rho -1| \leq \sigma$ and $|\varphi| \leq \sigma$ for the same $\sigma$ as earlier. 
Below, we only consider $q_s(\mathbf{x}, \omega)$ and $\widetilde{q}_s(\mathbf{x}, \omega)$ for $\mathbf{x} \in X$ far from all the receivers $\mathbf{x}_r$, and for $\omega \in \Omega$ with $\Omega$ sufficiently far from the origin (to avoid blowup of the Green's function $G(\mathbf{x}, \mathbf{x}_r, \omega)$).
Concretely, we assume 
\begin{equation}
\underset{\min_s{\substack{\mathbf{x} \in X\\ \omega \in \Omega}}}{\min} \left | q_s(\mathbf{x}, \omega) \right | \geq M 
\label{eq:M}
\end{equation}
for some universal number $M > 0$. 
Then, with $r_s(\mathbf{x}, \omega) e^{i \theta_s(\mathbf{x}, \omega)} = u_{0, s}(\mathbf{x}, \omega) \conj{q_s(\mathbf{x}, \omega)}$, we polar-decompose 
\begin{align}
    u_{0,s}(\mathbf{x}, \omega) \conj{\widetilde{q}_s(\mathbf{x}, \omega)} &= \widetilde{\alpha}_s(\mathbf{x}, \omega) e^{i\widetilde{\theta}_s(\mathbf{x}, \omega)} \nonumber \\
& = \alpha_s(\mathbf{x}, \omega) e^{i \theta_s (\mathbf{x}, \omega)} \rho(\mathbf{x}, \omega) e^{i \varphi(\mathbf{x}, \omega)} \nonumber \;.
\end{align}
From the size of $\varphi$, we have 
\begin{equation*}
    \left | \widetilde{\theta}_s(\mathbf{x}, \omega) - \theta_s(\mathbf{x}, \omega) \right | \leq \sigma\;.
\end{equation*}
Since $q_s(\mathbf{x}, \omega)$ and $\widetilde{q_s}(\mathbf{x}, \omega)$ are smooth in $\omega$ on a length scale of $\Delta \omega  = \nicefrac{2 \pi}{T}$, and further assuming that their amplitudes do not vanish, a similar result holds for their derivatives in $\omega$:
\begin{prop}
\label{prop:dt}
\begin{equation}
\left | \frac{\partial \widetilde{\theta}_s}{\partial \omega} (\mathbf{x}, \omega) - \frac{\partial \theta_s}{\partial \omega}(\mathbf{x}, \omega) \right | \leq c \sigma
\label{eq:dt}
\end{equation}
for some $c > 0$ that depends on the record length $T$, $G(\mathbf{x}, \mathbf{x}_r, \omega)$, the cutoff parameters $X$ and $\Omega$, and $M$ introduced earlier.
\end{prop}
The proof for proposition \ref{prop:dt} is in Appendix \ref{a:dt}.
\begin{remark}
From Proposition \ref{prop:dt}, it can be concluded that the resolution length scale\footnote{\label{fnr} $R = \frac{\left(\int||\mathbf{x}_0 - \mathbf{x}||I(\mathbf{x})^2d\mathbf{x}\right)^{1/2}}{\left ( \int I(\mathbf{x})^2 d\mathbf{x} \right )^{1/2}}$, where $\mathbf{x}$ is the spatial index, $\mathbf{x}_0$ is the known scatterer center, and $I(\mathbf{x})$ is the image value at that $\mathbf{x}$.
}, $R$, of the scatterer
is linearly dependent on the noise parameter, $\sigma$. 
This can be seen by examining the image of $I(\mathbf{x})$ of a single scatterer at $\mathbf{x} = 0$, following Equation \ref{eq:zic} where the phase mask is $\frac{\partial \theta}{\partial \omega} < \tau_{_{RES}}$.
Let $t$ be the distance, in units of time, of a putative scatterer that the imaging condition would place at $\mathbf{x}$; i.e., $t = || \mathbf{x}||/c_0$. In the noise-free case, $\frac{\partial \theta}{\partial \omega} = t$, so that the image can only be nonzero in a ball of radius $c_0\tau_{_{RES}}$ centered at the origin.
In the noisy case, $|\frac{\partial \theta}{\partial \omega}-t| \leq c\sigma$ for some $c>0$, so the condition $\frac{\partial \theta}{\partial \omega} < \tau_{_{RES}}$ means that
\begin{itemize}
\item the image will be computed in the classical manner (mask value = 1) when $t < \tau_{_{RES}} - c\sigma$
\item the image will be zero (mask value = 0) when $t > \tau_{_{RES}} + c\sigma$
\end{itemize}
This is illustrated in Figure \ref{fig:dtvt}.

\end{remark}

\begin{figure}
    \centering
    \includegraphics[width=0.75\textwidth]{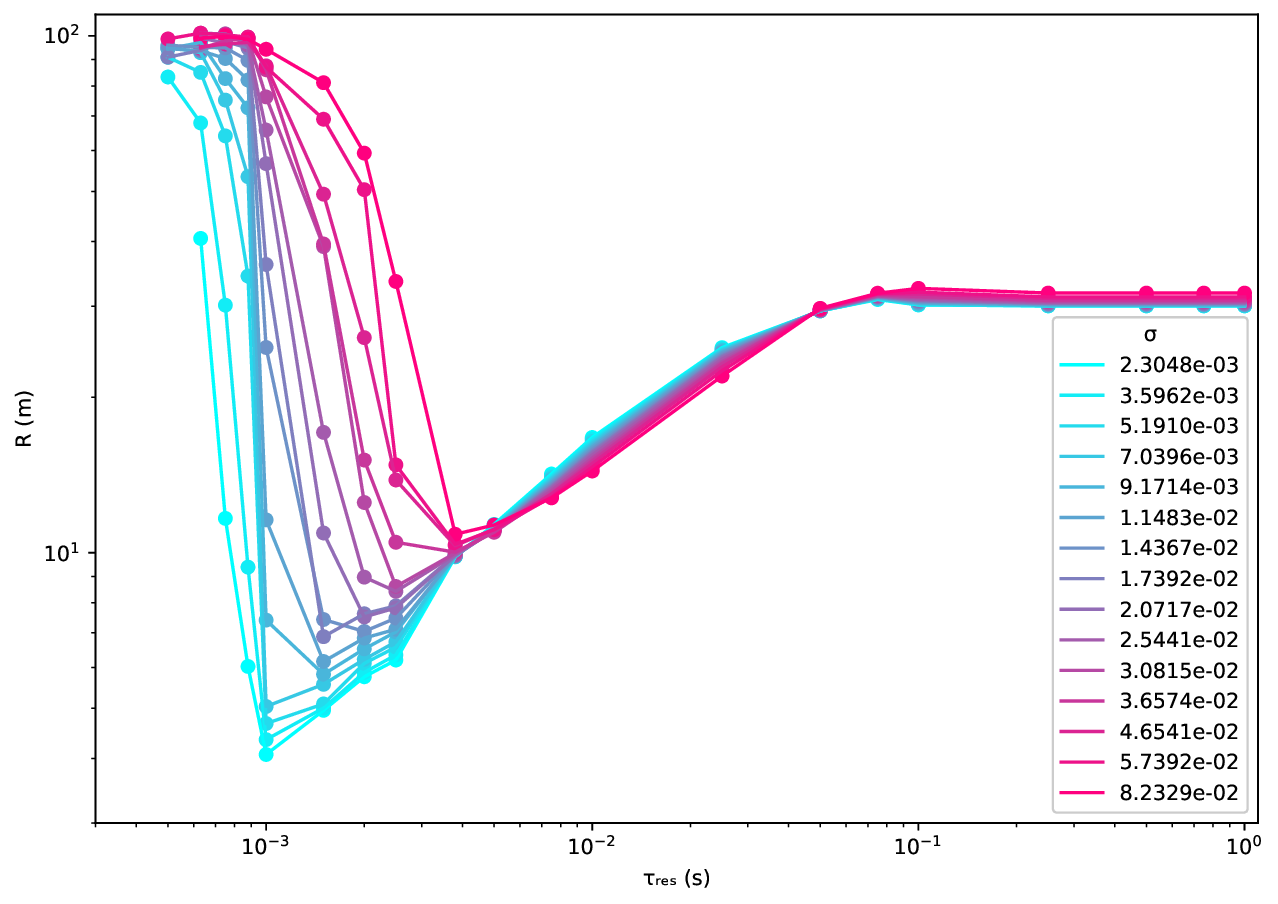}
     \caption{The mean resolved size of the scatterer, $R$, as a function of $\tau_{_{RES}}$ for various values of $\sigma$. }
     \label{fig:rvtau}
\end{figure}

\begin{figure}
    \centering
    \includegraphics[width=0.75\textwidth]{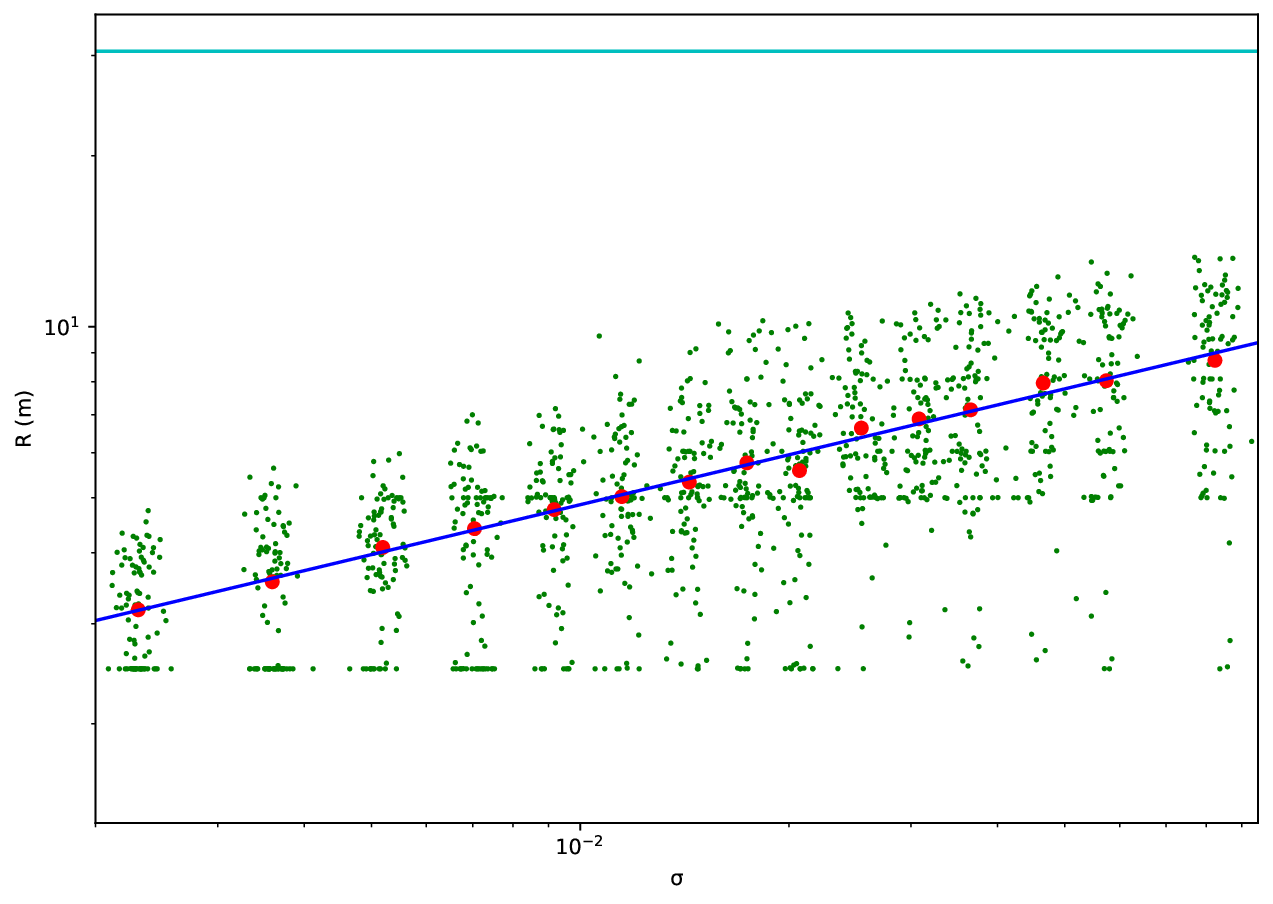}
     \caption{The resolved size of the scatterer as a function of $\sigma$ using the best $\tau_{_{RES}}$. Green points represent individual samples corresponding to different noise realizations, red points represent the mean for that noise value, and the blue line is fits the data with an $r^2$ value of $0.988$. The teal line represents the resolution with the standard imaging condition, which is calculated to be $0.273 \;\lambda$ (calculated from footnote \ref{fnr}), where $\lambda$ is the wavelength. The numerical grid spacing for the experiments was $2.5$ m, which limits the minimum resolvable size of the scatterer. }
     \label{fig:rvsigma}
\end{figure}

To demonstrate the performance of the method in the presence of noise, we perform numerical experiments. 
Here, we place a single point scatterer at the center of a $250 \times 250 $ m domain, with numerical grid spacing of $dz = dx = 2.5 $ m. 
Seven sources are evenly spread across the top surface of the domain, along with a receiver array with receivers placed $2.5 $ m apart. 
Fifteen values of $\sigma$ are chosen, which represent the standard deviation of Gaussian noise added to the data, as in Equation \ref{eq:noise}.
For each noise value, we run $90$ unique experiments and examine how resolved the scatterer is from various chosen values of $\tau_{_{RES}}$. 
We measure the resolution of the scatterer, $R$, as the radial extent from the center of the scatterer location (from footnote \ref{fnr}). 
The results are shown in Figure \ref{fig:rvtau}.
From this, we find the value of $\tau_{_{RES}}$ for each $\sigma$ that produces the minimum value of $R$---for a specific $\sigma$, images resolved using a smaller $\tau_{_{RES}}$ than this will be dominated by noise, while images with a larger $\tau_{_{RES}}$ will only be partially resolved.
This minimum value of $R$ vs its corresponding $\sigma$ is shown in Figure \ref{fig:rvsigma}. 
From this, it is clear that $R \sim \sigma$, as shown in Proposition \ref{prop:dt}.

Since the minimum resolvable size of the scatterer is limited by the numerical grid spacing, we only select noise values such that $R$ is larger than the $dz = dx = 2.5$ m grid spacing.

\subsection{Generalized zero-phase imaging condition}
\label{sec:gzpic}
The zero-phase imaging condition can also be generalized to an arbitrary model beyond a single scatterer with some careful modifications.
Ideally, we need the phase of the incident and adjoint fields to cancel out at any scattering location.
While the phase of the incident field remains the same as with a single scatterer, 
the phase of the adjoint field becomes corrupted due to the presence of multiple traveltimes attributed to each scattering point in the domain. 
At any given $\mathbf{x}$ location, we ideally want the adjoint field's phase to be solely attributed to that $\mathbf{x}$ location's traveltime. 
We can achieve this by careful windowing in the time domain---since the adjoint field only contributes to the imaging condition when it coincides in space and time with the forward field, we can create a new modified adjoint field where the amplitude is smoothly set to zero where it does not coincide with the incident field.
That way, in the Fourier domain, the only phase information for the modified adjoint field at a given $\mathbf{x}$ location is the traveltime to that specific location.

We choose to do this in the Fourier domain by convolving the adjoint field with a Gaussian function propagating with the incident field's arrival time.
We can define a new modified adjoint field, $q_{s, \text{mod}}(\mathbf{x}, \omega)$, such that 
\begin{equation}
q_{s, \text{mod}}(\mathbf{x}, \omega) = q_s(\mathbf{x}, \omega) *  \left[ e^{i \omega \tau_{0,s}(\mathbf{x})} g(\omega) \right] \;,
\end{equation}
where $*$ is convolution, $g(\omega)$ is a Gaussian function of tunable width, and $\tau_{0,s}(\mathbf{x})$ is the arrival time of the incident field, which can be found as $\tau_{0,s}(\mathbf{x}) = \argmax_t u_{0, s}(\mathbf{x}, t)$. 
If this estimation of $\tau_{0, s}(\mathbf{x})$ is exact, then it is $\tau(\mathbf{x}_s, \mathbf{x})$.

\subsection{Limitations}
\label{sec:limitations}

Using the modification in Section \ref{sec:gzpic},
we can expect to achieve some superresolution in cases outside of a single scattering point. 
This should be able to perfectly superresolve models in which $q_{s, \text{mod}}(\mathbf{x}, \omega)$ only has contributions from one scattering point at any given $\mathbf{x}$ (e.g., where scatterers are located sufficiently far from each other). 
However, in models where $q_{s, \text{mod}}(\mathbf{x}, \omega)$ still has traveltime contributions from multiple scattering points (e.g., Figure \ref{fig:wavy}), the resultant phase at any given $\mathbf{x}$ is slightly corrupted from nearby scattering contributions. Thus, the $\Gamma(\cdot)$ operator may not perfectly resolve scattering point locations. 
In this case, using too small of a $\tau_{_{RES}}$ may erroneously remove sections of the image. 
To alleviate this issue, at the expense of not achieving full superresolution, one may select a larger $\tau_{_{RES}}$ while still allowing for improvements in resolution compared to results from the standard imaging condition.
As with many forms of superresolution, this method is inherently delicate and cannot be expected to perform on complex models with intricate phase contributions.
Standard community models such as Marmousi are difficult to handle in our framework due to the overwhelming presence of unfavorable phase combinations.

Additionally, in more complex geometry, artifacts may result due to the affects of multipathing. 
When this occurs, the forward and adjoint wavefields may consist of a superposition of wavefronts,
which corrupts the phase by the presence of more than one traveltime. 
As a result, this will produce artifacts which cannot be resolved by the standard or zero-phase imaging condition.

\section{Numerical Experiments}

\subsection{Single scatterer}

We place a scatterer of size $10\; m \times 10\; m$ with velocity of $3500\; \nicefrac{m}{s}$ in a background medium with a velocity of $2800\; \nicefrac{m}{s}$. 
The domain is of size $2000 \;m \times 2000 \;m$, with grid spacing $dz = dx = 10 \; m$. 
We place a receiver array on the top surface of the domain, with receivers spaced $10 \; m$ apart, and have seven $25\; Hz$ Ricker\footnote{\label{fn2} second derivative of a Gaussian} wavelet sources spread evenly across the top surface.
We forward model with the constant density acoustic wave equation without any additional noise.

\begin{figure}
     \centering
     \begin{subfigure}[b]{0.32\textwidth}
         \centering
         \includegraphics[width=\textwidth]{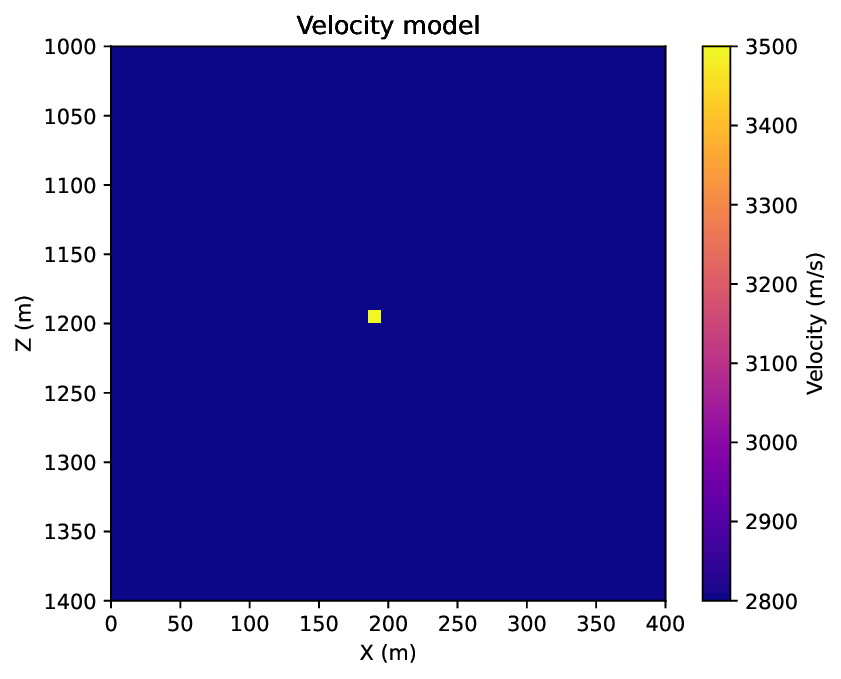}
     \end{subfigure}
     \hfill
     \begin{subfigure}[b]{0.32\textwidth}
         \centering
         \includegraphics[width=\textwidth]{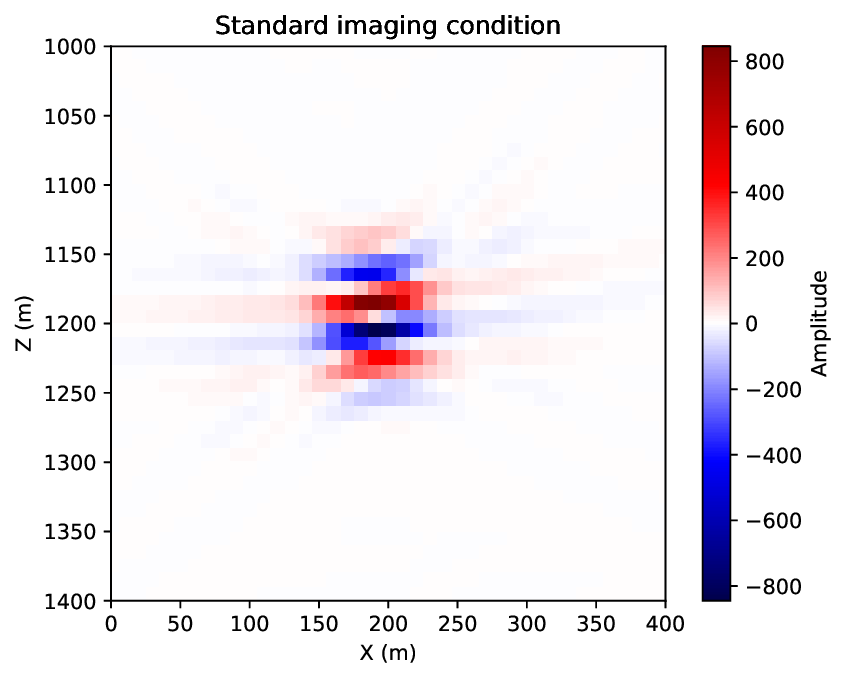}
     \end{subfigure}
     \hfill
     \begin{subfigure}[b]{0.32\textwidth}
         \centering
         \includegraphics[width=\textwidth]{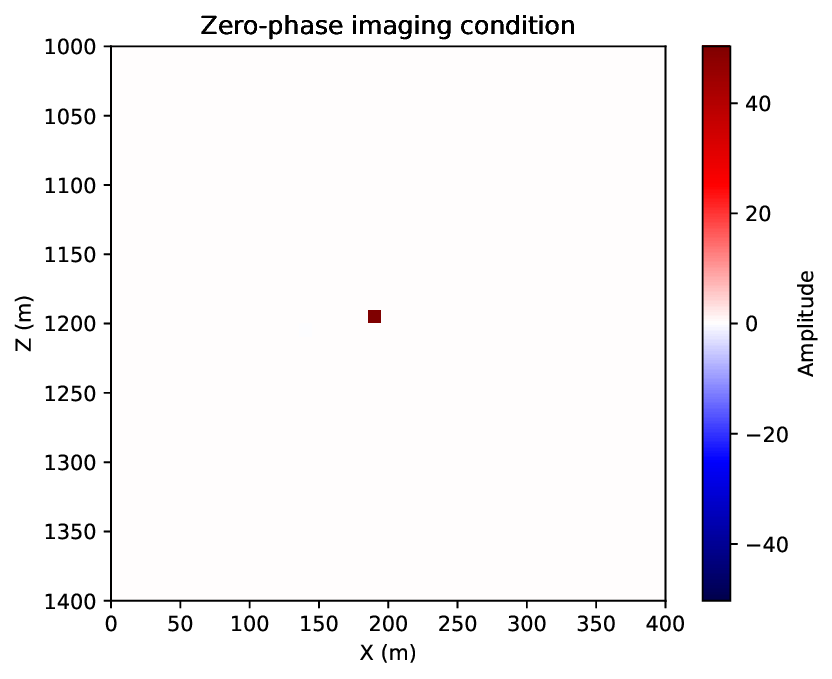}
     \end{subfigure}
     \caption{A velocity model with a scatterer located at $(200\; m, 1200\; m)$ (left), an image produced with the standard imaging condition (center), and an image produced with the zero-phase imaging condition (right). }
     \label{fig:wavy}
\end{figure}

\subsection{Two wavy reflectors: zero-mean oscillatory interfaces} \label{sec:2wr}
\begin{figure}
     \centering
     \begin{subfigure}[b]{0.48\textwidth}
         \centering
         \includegraphics[width=\textwidth]{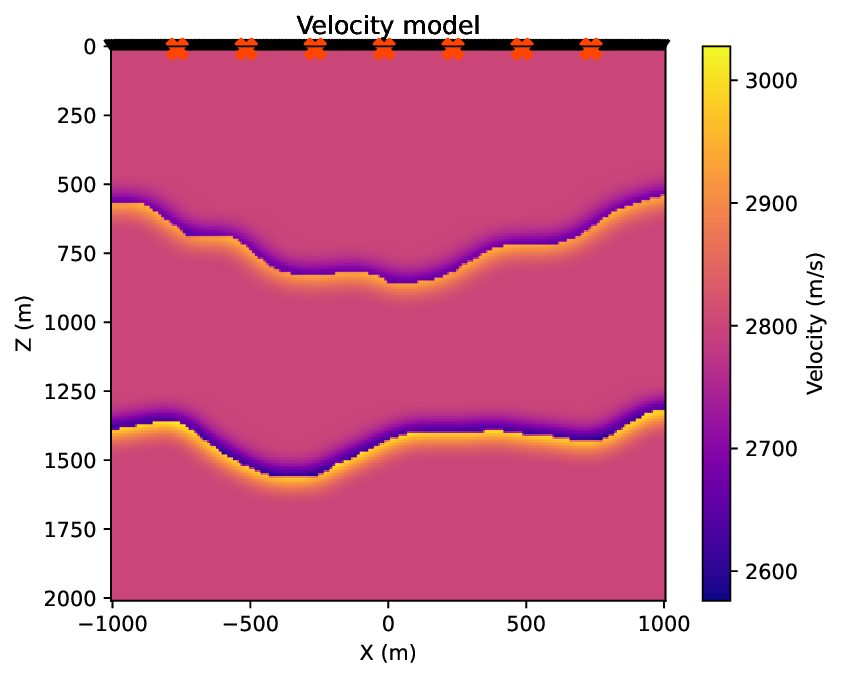}
         \caption{}
         \label{fig:2wvel}
     \end{subfigure}
     \hfill
     \begin{subfigure}[b]{0.48\textwidth}
         \centering
         \includegraphics[width=\textwidth]{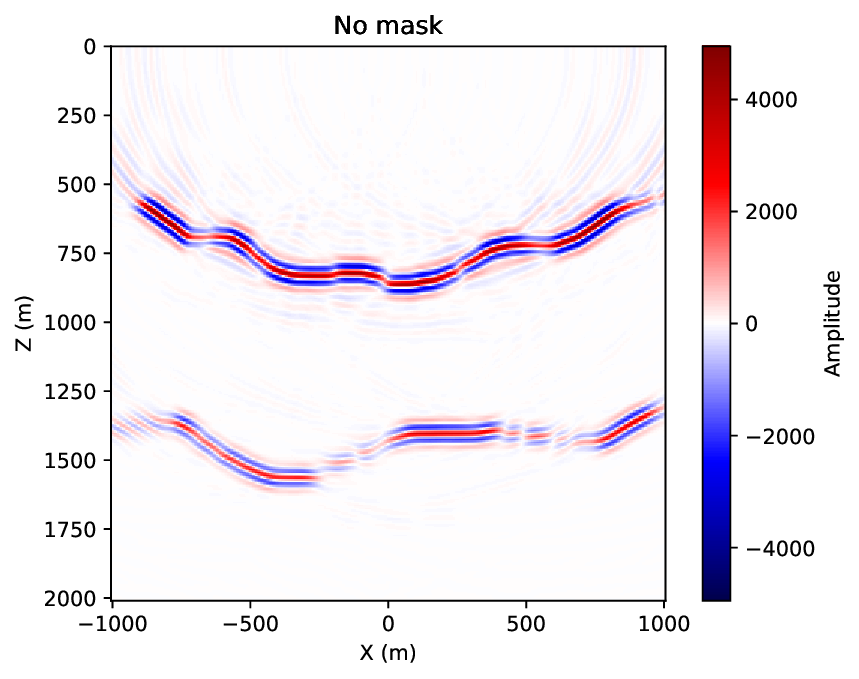}
         \caption{}
         \label{fig:2wrtm}
     \end{subfigure}
     \hfill
     \begin{subfigure}[b]{0.48\textwidth}
         \centering
         \includegraphics[width=\textwidth]{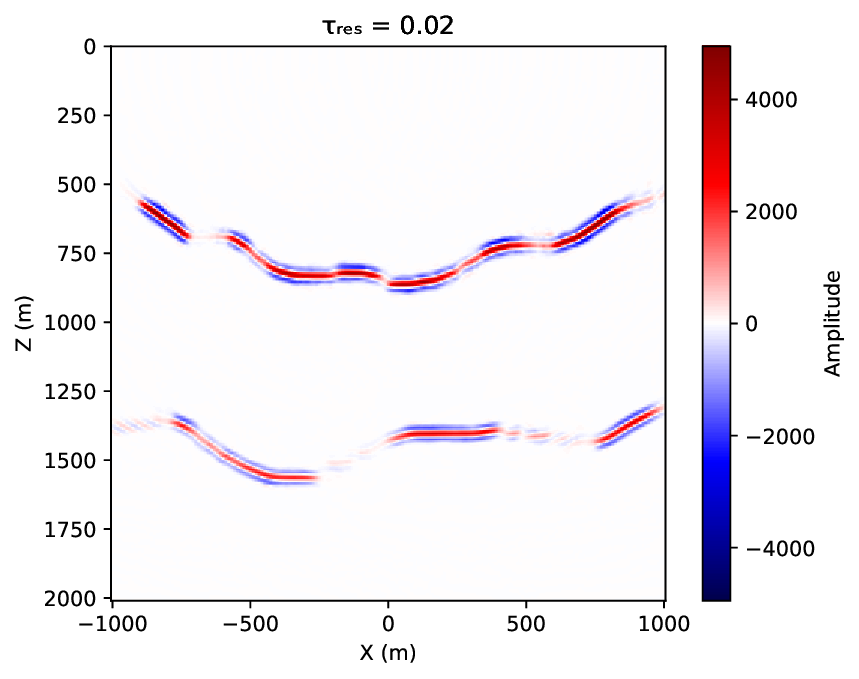}
         \caption{}
         \label{fig:2w0.02}
     \end{subfigure}
     \hfill
     \begin{subfigure}[b]{0.48\textwidth}
         \centering
         \includegraphics[width=\textwidth]{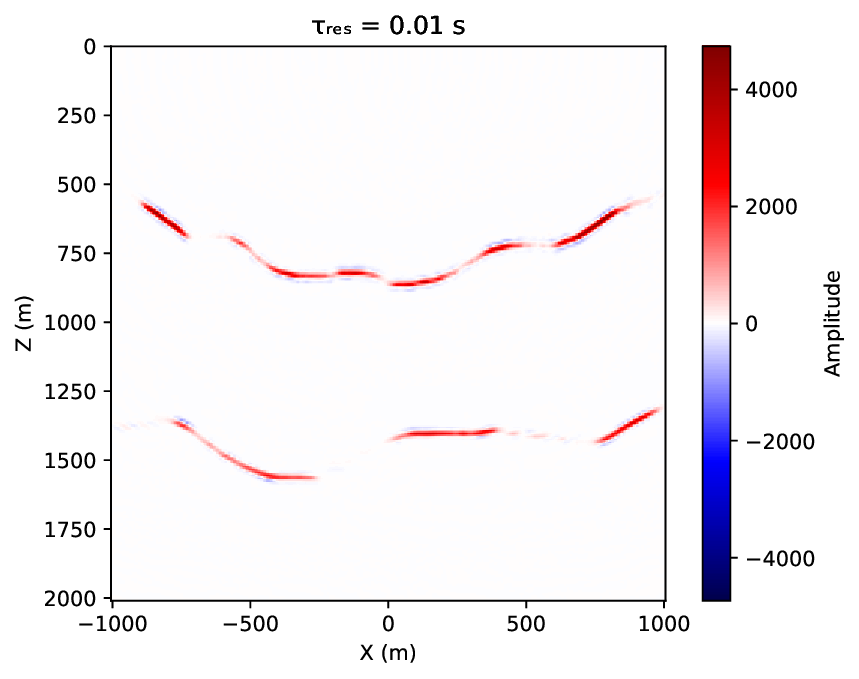}
         \caption{}
         \label{fig:2w0.01}
     \end{subfigure}
     \caption{A more complex example using the zero-phase imaging condition using zero-mean oscillatory reflection interfaces. (a) ground truth velocity model; (b) resultant image using the standard imaging condition; (c) resultant image using the zero-phase imaging condition with $\tau_{_{RES}} = 0.02\;s$; (d) resultant image using the zero-phase imaging condition with $\tau_{_{RES}} = 0.01\;s$.}
     \label{fig:wavy}
\end{figure}

Since our method relies on the forward and adjoint fields coinciding exactly in space and time at discontinuities in the domain, we require that our velocity model have zero-mean perturbation oscillatory reflectors. 
Without this, the forward and adjoint fields will have different arrival times, and thus won't produce kinematically correct results with this method.
We choose an example velocity model with two wavy zero-mean perturbation reflectors, as shown in Figure \ref{fig:2wvel}.
Our domain is $2 \; km \times 2 \; km$, with a grid spacing of $dz = dx = 10\;m$, and use a $25 \; Hz$ Ricker\footnoteref{fn2} wavelet as our source with $dt = 0.001 \;s$ with a record length of $2.5 \;s$.
We place an array of receivers on the surface of the domain, spaced $10 \; m$ apart, and spread 7 sources equally across the surface.
The results of reverse-time migration (RTM) using the standard imaging condition are shown in Figure \ref{fig:2wrtm}. 
We use the zero-phase imaging condition with $\tau_{_{RES}} = 0.02 \;s$ in Figure \ref{fig:2w0.02} and $\tau_{_{RES}} = 0.01 \;s$ in Figure \ref{fig:2w0.01}.
The chosen value of $\tau_{_{RES}}$ indicates the amount the reflectors can be resolved.

The gaps in the resultant images where there should be energy are due to interference in the phase of $q_{s, \text{mod}}(\mathbf{x}, \omega)$ from the presence of multiple nearby scattering points. Thus, the $\Gamma(\cdot)$ operator, when selected with a low enough $\tau_{_{RES}}$, may filter out true scattering point locations.
This is justified further in section \ref{sec:limitations}.

\subsection{Two wavy reflectors: non-zero-mean oscillatory interfaces}
\begin{figure}
     \centering
     \begin{subfigure}[b]{0.48\textwidth}
         \centering
         \includegraphics[width=\textwidth]{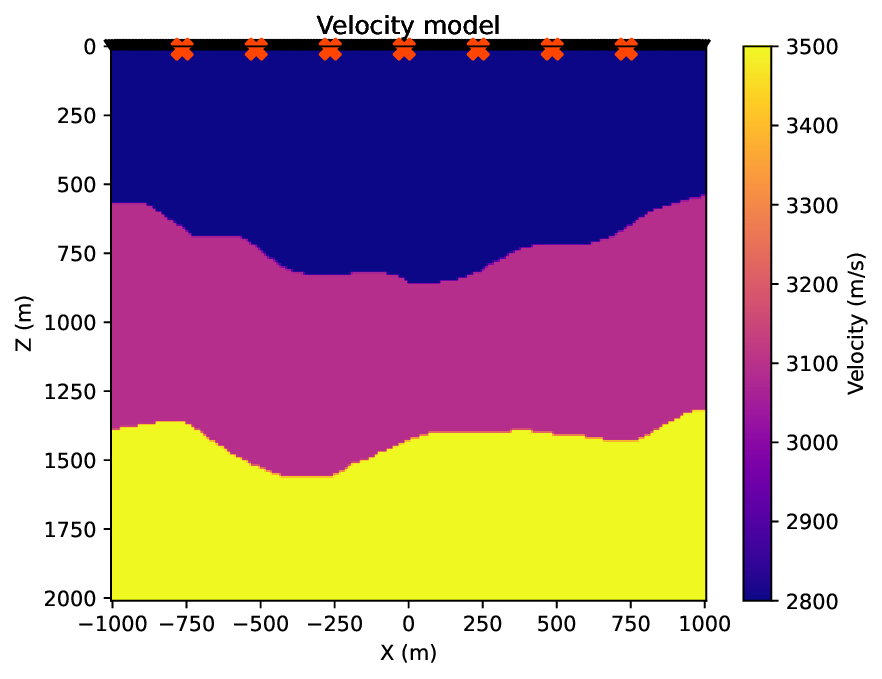}
         \caption{}
         \label{fig:2wvel2}
     \end{subfigure}
     \hfill
     \begin{subfigure}[b]{0.48\textwidth}
         \centering
         \includegraphics[width=\textwidth]{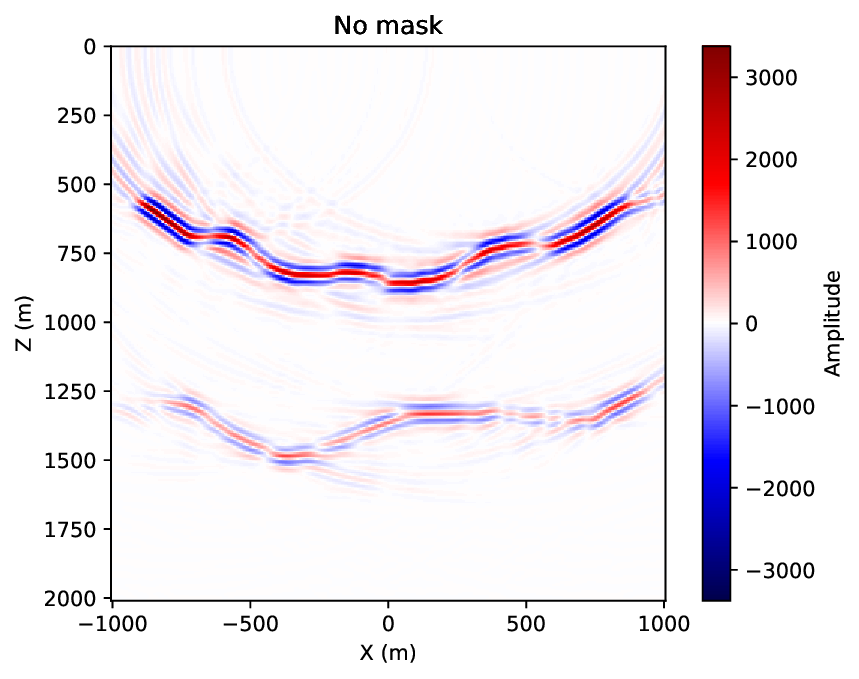}
         \caption{}
         \label{fig:2wrtm2}
     \end{subfigure}
     \hfill
     \begin{subfigure}[b]{0.48\textwidth}
         \centering
         \includegraphics[width=\textwidth]{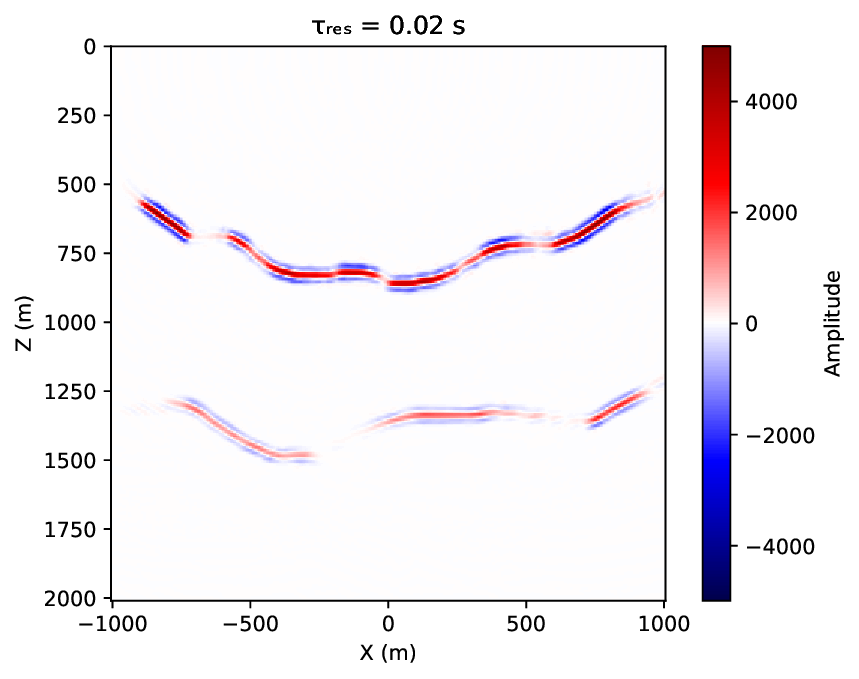}
         \caption{}
         \label{fig:2w0.022}
     \end{subfigure}
     \hfill
     \begin{subfigure}[b]{0.48\textwidth}
         \centering
         \includegraphics[width=\textwidth]{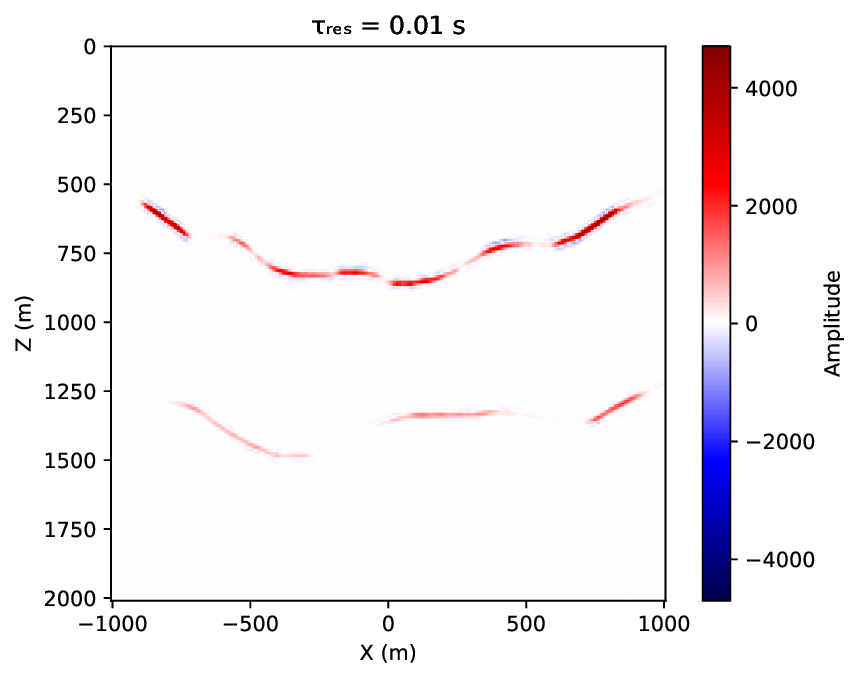}
         \caption{}
         \label{fig:2w0.012}
     \end{subfigure}
     \caption{A more complex example using the zero-phase imaging condition using non-zero-mean oscillatory reflection interfaces. (a) ground truth velocity model; (b) resultant image using the standard imaging condition; (c) resultant image using the zero-phase imaging condition with $\tau_{_{RES}} = 0.02\;s$; (d) resultant image using the zero-phase imaging condition with $\tau_{_{RES}} = 0.01\;s$. This is comparable to Figure \ref{fig:wavy}.}
     \label{fig:wavy2}
\end{figure}

Although zero-mean perturbation reflection interfaces are theoretically required for our method, the zero-phase imaging condition may still work efficiently in less favorable models. To test this, we extend the example from subsection \ref{sec:2wr}, shown in Figure \ref{fig:wavy}, to a non-zero-mean example. In this example, we use the same parameters from subsection \ref{sec:2wr}, except we replace the velocity model from Figure \ref{fig:2wvel}, which has zero-mean oscillatory reflection interfaces, for the one in Figure \ref{fig:2wvel2}, which is the same model with non-zero-mean reflection interfaces. The results are shown in Figure \ref{fig:wavy2}. Despite having theoretically unfavorable phase contributions, the results from this example are nearly comparable to that from \ref{fig:wavy}, with the notable exception that the resultant images are more poorly resolved with depth.

\begin{figure}
     \centering
     \begin{subfigure}[b]{0.33\textwidth}
         \centering
         \includegraphics[width=\textwidth]{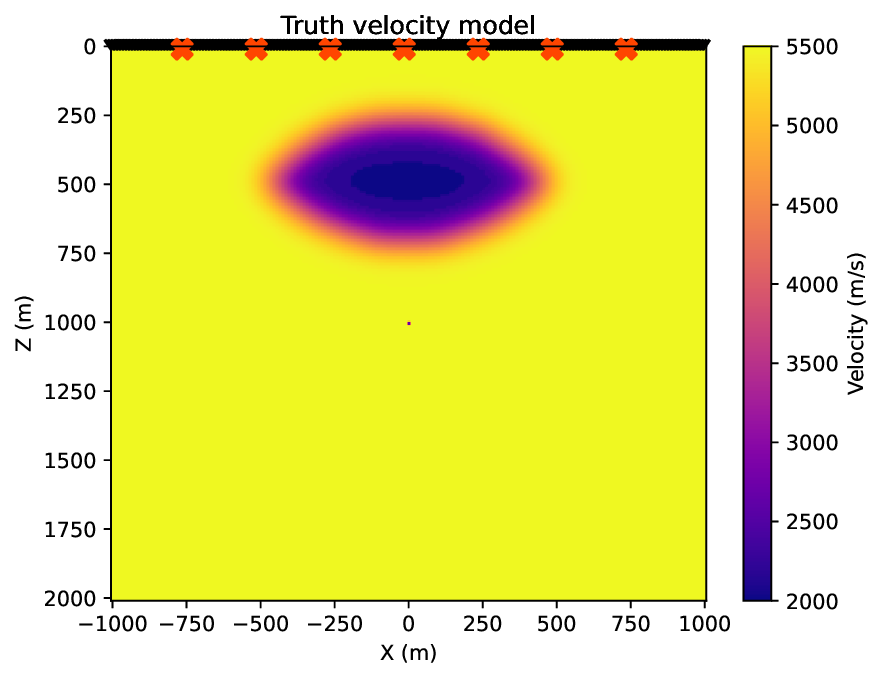}
         \caption{}
         \label{fig:lwvel2}
     \end{subfigure}
     \hfill
     \begin{subfigure}[b]{0.315\textwidth}
         \centering
         \includegraphics[width=\textwidth]{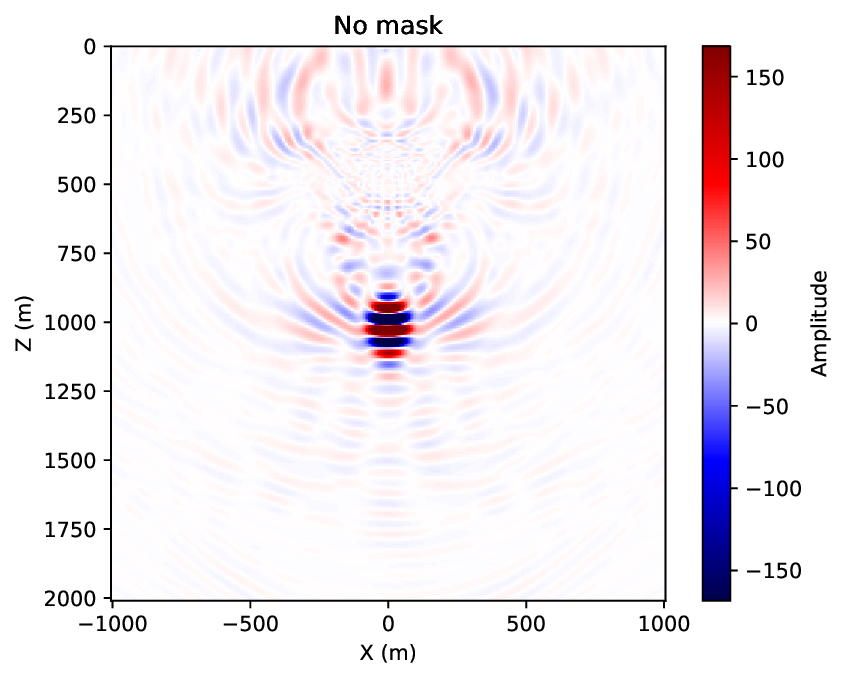}
         \caption{}
         \label{fig:lw0.022}
     \end{subfigure}
     \hfill
     \begin{subfigure}[b]{0.315\textwidth}
         \centering
         \includegraphics[width=\textwidth]{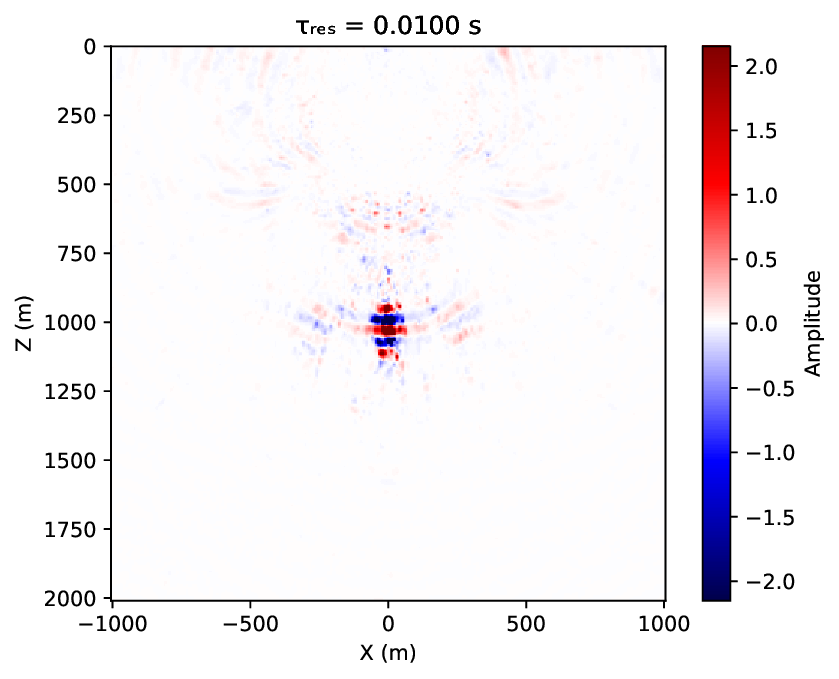}
         \caption{}
         \label{fig:lw0.012}
     \end{subfigure}
     \caption{An example introducing caustics into the data. (a) ground truth velocity model, where the migration velocity model includes the lens without the scatterer located at (1000, 0); (b) resultant image using the standard imaging condition; (c) resultant image using the zero-phase imaging condition with $\tau_{_{RES}} = 0.01$ s.}
     \label{fig:lens}
\end{figure}

\subsection{Acoustic lens}
In a final example, we introduce caustics into the data by adding an acoustic lens located above a scatterer we are interested in imaging. This numerical example is demonstrated in Figure \ref{fig:lens}. Using the standard imaging condition, the resultant image (Figure \ref{fig:lw0.022}) is a blurred representation of the scatterer. As discussed in Section \ref{sec:limitations}, the zero-phase imaging condition should not provide better results than the standard imaging condition due to the presence of multiple traveltimes attributed to the same scattering point. 
This expected result is demonstrated in Figure \ref{fig:lw0.012}. 

\section{Conclusion}
In this paper, we describe a method of superresolution for simple imaging applications, which is performed at the level of the imaging condition defined by the adjoint-state method. 
In the presence of noise, for a single scatterer, the achieved resolution level is proportional to the noise level.

\section{Acknowledgements}
SG acknowledges support from the United States Department of Energy through the Computational Science Graduate Fellowship (DOE CSGF) under grant number DE-SC0019323.
LD is also supported by AFOSR grant FA9550-17-1-031.

\appendix

\makeatletter
\renewcommand{\@seccntformat}[1]{\csname the#1\endcsname\quad}
\makeatother
\renewcommand{\thesection}{\Alph{section}}
\numberwithin{equation}{section}

\section{Stationary phase points}
\label{ap:stphpt}
We wish to find the locations of stationary phase of the $r$-dependent adjoint field; that is, where $\mathbf{t}(\mathbf{x}_r) \cdot \nabla_{\mathbf{x}_r} \phi (\mathbf{x}_r) = 0$.
\begin{prop}
\label{p:dir}
Let $\mathbf{\xi} = \nabla_{\mathbf{x}} \tau(\mathbf{x}, \mathbf{y})$. Then $\mathbf{a}\cdot \nabla_{\mathbf{x}} \tau(\mathbf{x}, \mathbf{y}) = (\mathbf{a} \cdot \hat{\mathbf{\xi}}) | \nabla_{\mathbf{x}}\tau(\mathbf{x}, \mathbf{y})|$, where $\hat{\mathbf{\xi}} = \nicefrac{\mathbf{\xi}}{|\mathbf{\xi}|}$.
\end{prop}
\begin{proof}
\begin{equation*}
\mathbf{a} = (\mathbf{a} \cdot \hat{\mathbf{\xi}})\hat{\mathbf{\xi}} + (\mathbf{a} \cdot \hat{\mathbf{\xi}}^{\perp})\hat{\mathbf{\xi}}^{\perp}
\end{equation*}
Then 
\begin{align}
\mathbf{a}\cdot \nabla_{\mathbf{x}}\tau(\mathbf{x}, \mathbf{y}) = \mathbf{a}\cdot \mathbf{\xi} &= (\mathbf{a} \cdot \hat{\mathbf{\xi}})\hat{\mathbf{\xi}}\cdot\mathbf{\xi} + 0 \nonumber \\
&= (\mathbf{a} \cdot \mathbf{\xi})|\mathbf{\xi}| \nonumber \\
&= (\mathbf{a} \cdot \hat{\xi}) |\nabla_{\mathbf{x}}\tau(\mathbf{x}, \mathbf{y})| \nonumber\;.
\end{align}
\end{proof}
From Proposition \ref{p:dir}, with $\mathbf{\xi}(\mathbf{y}) = \nabla_{\mathbf{x}_r} \tau(\mathbf{y}, \mathbf{x}_r)$, the tangential derivative of $\phi(\mathbf{x}_r)$ is then  
\begin{align}
\mathbf{t}(\mathbf{x}_r) \cdot \nabla_{\mathbf{x}_r} \phi(\mathbf{x}_r) & = \left (\mathbf{t}(\mathbf{x}_r)\cdot \hat{\mathbf{\xi}}(\mathbf{x}^*)\right ) \left | \nabla_{\mathbf{x}_r}\tau(\mathbf{x}^*, \mathbf{x}_r)\right | 
 - \left (\mathbf{t}(\mathbf{x}_r)\cdot \hat{\mathbf{\xi}}(\mathbf{x})\right ) \left | \nabla_{\mathbf{x}_r}\tau(\mathbf{x}_r, \mathbf{x})\right |  \nonumber \\
 &= \frac{1}{c(\mathbf{x}_r)} \left [ \mathbf{t}(\mathbf{x}_r) \cdot \hat{\mathbf{\xi}}(\mathbf{x}^*) - \mathbf{t}(\mathbf{x}_r) \cdot \hat{\mathbf{\xi}}(\mathbf{x})\right ] \;,
\label{eq:td}
\end{align}
since $| \nabla_{\mathbf{x}_r} \tau(\mathbf{x}^*, \mathbf{x}_r) | =  | \nabla_{\mathbf{x}_r} \tau(\mathbf{x}_r, \mathbf{x}) | = \nicefrac{1}{c(\mathbf{x}_r)}$ from the eikonal equation.   
Then Equation \ref{eq:td} equals zero in the case when $\hat{\mathbf{\xi}}(\mathbf{x}^*) = \hat{\mathbf{\xi}}(\mathbf{x})$, which is when both $\hat{\mathbf{\xi}}$ rays are identical.
Conversely, if $\mathbf{t}(\mathbf{x}_r) \cdot \nabla_{\mathbf{x}_r}\phi(\mathbf{x}_r) = 0$, then $\frac{1}{c(\mathbf{x}_r)} \left [ \mathbf{t}(\mathbf{x}_r) \cdot \hat{\mathbf{\xi}}(\mathbf{x}^*) - \mathbf{t}(\mathbf{x}_r) \cdot \hat{\mathbf{\xi}}(\mathbf{x}) \right ] = 0$.
This occurs when $\mathbf{x}_r$ is on the ray passing through $\left [ \mathbf{x}, \mathbf{x}^* \right ]$. 

\section{Locations of zero phase}
\label{a:antipodal}
\begin{figure}
\centering
\includegraphics{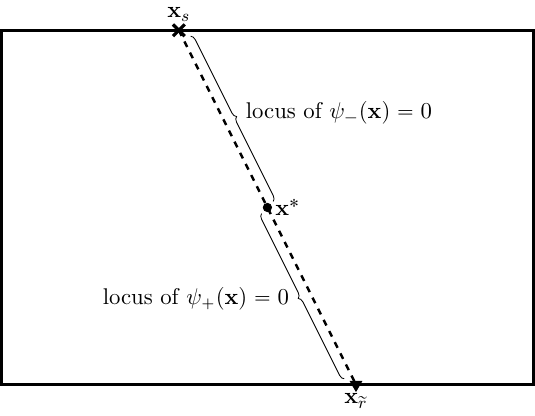}
%
%
%
%
%
%

\caption{The ray through a source located at $\mathbf{x}_s$ and a scatterer located at $\mathbf{x}^*$ indicates where the phase of the integrand of the imaging condtion (Equation \ref{eq:pic}) will be zero. 
With a special receiver $\mathbf{x}_{\widetilde{r}}$ antipodal to a given source location $\mathbf{x}_s$, the locus of zero phase when $\psi_+ (\mathbf{x})= 0$ is when $\mathbf{x} \in [\mathbf{x}^*, \mathbf{x}_{\widetilde{r}}]$, and the locus of zero phase for when $\psi_- (\mathbf{x})= 0$ is when $\mathbf{x}^* \in [\mathbf{x}_s, \mathbf{x}^*]$. 
Therefore, we restrict our receiver array such there is no receiver antipodal to any source location through the scatterer location to avoid including this transmission artifact in the zero-phase imaging condition.
}
\label{fig:icpm}
\end{figure}

We are interested in finding the $\mathbf{x}$ locations where the phase of the integrand of the imaging condition (Equation \ref{eq:pic}) is zero, dependent on source location $\mathbf{x}_s$, scatterer location $\mathbf{x}^*$, and the 
receiver location $\mathbf{x}_{\widetilde{r}}$.
We split Equation \ref{eq:pic} into two cases: when $\psi_{+}(\mathbf{x}) = 0$ and when $\psi_{-}(\mathbf{x}) = 0$.
In both cases, a special receiver location $\mathbf{x}_{\widetilde{r}}$ is antipodal to the source location $\mathbf{x}_s$ from the scatterer location $\mathbf{x}^*$ from the analysis from Equation \ref{eq:phi}.
In the case of $\psi_{+}(\mathbf{x}) = 0$, 
$\psi_{+}(\mathbf{x}) = \tau(\mathbf{x}_s, \mathbf{x}) - \tau(\mathbf{x}_s, \mathbf{x}^*) - \tau(\mathbf{x}^*, \mathbf{x}) = 0$ only holds true when $\mathbf{x}^*$ is on the $[\mathbf{x}, \mathbf{x}_s]$ ray. 
Thus, $\psi_{+}(\mathbf{x}) = 0$ when $\mathbf{x} \in [\mathbf{x}^*, \mathbf{x}_{\widetilde{r}}]$.
In the case of $\psi_{-}(\mathbf{x}) = 0$, 
$\psi_{-}(\mathbf{x}) = \tau(\mathbf{x}_s, \mathbf{x}) - \tau(\mathbf{x}_s, \mathbf{x}^*) + \tau(\mathbf{x}^*, \mathbf{x}) = 0$ only holds true when $\mathbf{x}$ is on the $[\mathbf{x}_s, \mathbf{x}^*]$ ray. 
Thus, $\psi_{-}(\mathbf{x}) = 0$ when $\mathbf{x} \in [\mathbf{x}_s, \mathbf{x}^*]$.
The locus of zero phase for both cases is illustrated in Figure \ref{fig:icpm}.

In the union of $\psi_{+}(\mathbf{x})$ and $\psi_{-}(\mathbf{x})$, $\psi_{\pm}(\mathbf{x}) = 0$ on the whole $[\mathbf{x}_s, \mathbf{x}_{\widetilde{r}}]$ ray though $\mathbf{x}^*$. 
This transmission artifact occurs when any $\mathbf{x}_{r}$ are antipodal to any $\mathbf{x}_s$. 
Henceforth, we always assume $\mathbf{x}_r$ are not antipodal to any $\mathbf{x}_s$. 

\section{Derivative of phase with noise}
\label{a:dt}
We wish to prove Proposition \ref{prop:dt} by showing that $\frac{\partial}{\partial \omega} (\widetilde{\theta} - \theta)$ is at most proportional to $\sigma$. 
We have our adjoint field $q(\mathbf{x},\omega) = r e^{i \theta}$, our noisy adjoint field $\widetilde{q}(\mathbf{x},\omega) = \widetilde{r} e^{i \widetilde{\theta}}$, $r \neq 0$, $\widetilde{r} \neq 0$, $\varphi = \widetilde{\theta} - \theta$, and $q(\mathbf{x}, \omega)$ and $\widetilde{q}(\mathbf{x},\omega)$ are smooth in in $\omega$.
From Equation \ref{eq:mulmod}, $\varphi = \operatorname{Im}( \log ( \nicefrac{\widetilde{q}}{q}))$, then, with $\cdot '$ denoting $\nicefrac{\partial}{\partial \omega}$,
\begin{align}
\nonumber
\varphi' & = \operatorname{Im} \left [ \frac{\nicefrac{\widetilde{q}'}{q}'}{\nicefrac{\widetilde{q}}{q}}\right ] 
= \frac{r}{\widetilde{r}} \frac{\widetilde{q}'}{q'}
= \frac{r}{\widetilde{r}} \left [ \frac{\widetilde{q}'}{q} - \frac{q'}{q} \frac{\widetilde{q}}{q}\right ]
= \frac{e^{-i \theta}}{\widetilde{r}} \widetilde{q}' - \frac{1}{r} e^{i (\widetilde{\theta} - 2\theta)}q' \\
\nonumber & = e^{-i \theta} \left ( \frac{\widetilde{q}'}{\widetilde{r}} - \frac{q'}{r}e^{i\varphi} \right )  \\
\nonumber & = e^{-i \theta} \left ( \frac{\widetilde{q}' - q'}{\widetilde{r}} + q' \left ( \frac{1}{\widetilde{r}} - \frac{1}{r}e^{i\varphi} \right ) \right )  \\
& = e^{-i \theta} \left ( \frac{\widetilde{q}' - q'}{\widetilde{r}} + q' \left ( \frac{r - \widetilde{r}}{r\widetilde{r}} \right ) + \frac{q'}{r} \left ( 1 - e^{i\varphi} \right ) \right )  \label{eq:dpdw}\;.
\end{align}
Additionally,  
\begin{align}
\nonumber \left | \frac{\partial \widetilde{q}}{\partial \omega} - \frac{\partial q}{\partial \omega} \right | & = \left | \frac{\partial}{\partial \omega} \int_0^T e^{-i\omega t} \left [ \widecheck q (\mathbf{x}, t) - \widecheck{\widetilde{q}}(\mathbf{x}, t)\right ]  \right | \\ 
\nonumber & = \left | \int_0^T (-it)e^{-i\omega t} \left ( \widecheck{q}(\mathbf{x}, t) - \widecheck{\widetilde{q}}(\mathbf{x},t) \right )dt\right |\\
\nonumber & \leq T \int_0^T \left |  \widecheck{q}(\mathbf{x}, t) - \widecheck{\widetilde{q}}(\mathbf{x},t) \right | dt\\
\nonumber & \leq T^2 \max_t \left | \widecheck{q}(\mathbf{x}, t) - \widecheck{\widetilde{q}}(\mathbf{x}, t) \right | dt \\
\nonumber & \leq T^2 \max_t \left | \int_{\Omega} e^{i\omega t} \left ( q(\mathbf{x}, \omega) - \widetilde{q}(\mathbf{x}, \omega) \right ) \frac{d\omega}{2 \pi}\right | \\
\nonumber & \leq \frac{T^2}{2\pi}\left [ \underset{{\substack{\mathbf{x} \in X\\ \omega \in \Omega}}}{\max} \sum_r \left | G(\mathbf{x}, \mathbf{x}_r, \omega) \right | \right ] \max_r \int_{\Omega}\left |  n_{r, s, \omega} \right | d\omega  \qquad \text{\emph{(From Equation \ref{eq:paf})}} \\
& \leq D \sigma \label{eq:dqdw}\;,
\end{align}
where $D = \frac{T^2}{2\pi}\left [ \underset{{\substack{\mathbf{x} \in X\\ \omega \in \Omega}}}{\max} \sum_r \left | G(\mathbf{x}, \mathbf{x}_r, \omega) \right | \right ] \int_\Omega \left | d_{r, \omega, s}\right | d\omega$, and $\widecheck{\cdot}$ denotes the inverse Fourier transform. From $|\frac{r - \widetilde{r}}{r}| = | 1 - \rho | \leq \sigma$, $|1 - e^{i \varphi}| \leq | \varphi| \leq \sigma$, and Equations \ref{eq:dpdw} and \ref{eq:dqdw},
\begin{align}
\left | \frac{\partial}{\partial \omega} \varphi \right | & \leq \left | \frac{1}{\widetilde{r}} \left ( \widetilde{q}' - q'\right ) \right | + \left | \frac{q'}{\widetilde{r}} \left ( \frac{r - \widetilde{r}}{r}\right )\right | + \left | \frac{q'}{r} \left ( 1 - e^{i\varphi} \right ) \right |\nonumber\\
& \leq \left | \frac{1}{\widetilde{r}} D \sigma \right | + \left | \frac{q'}{\widetilde{r}} \sigma \right | + \left | \frac{q'}{r} \sigma \right | \nonumber  \\
& \leq \left ( 2\frac{D}{M} + 2\frac{D}{M} + \frac{D}{M} \right ) \sigma = 5 \frac{D}{M} \sigma\;,
\end{align}
where $M$ is from Equation \ref{eq:M}, and, using analagous reasoning to the result in Equation \ref{eq:dqdw}, it can be seen that $\underset{{\substack{\mathbf{x} \in X\\ \omega \in \Omega}}}{\max} \left | q_s'(\mathbf{x}, \omega) \right | \leq D $.

\printbibliography
\end{document}